\documentstyle[12pt]{article}

        {\newpage
        \setcounter{page}{1}}

\renewcommand{\title}[1]{%
        {\begin{center}
        \Large\bf #1
        \end{center}}
        \vskip .3in}

\renewcommand{\author}[1]{%
        {\begin{center}
        #1
        \end{center}}}

\renewcommand{\abstract}[1]{%
        \begin{center}%
        {\vspace{1em}\vspace{0pt}\bf Abstract}%
        \end{center}%
        \noindent #1}

\renewcommand{\date}[1]{%
        \begin{center}%
        #1%
        \end{center}}

\newcommand{\drawsquare}[2]{\hbox{%
\rule{#2pt}{#1pt}\hskip-#2pt
\rule{#1pt}{#2pt}\hskip-#1pt
\rule[#1pt]{#1pt}{#2pt}}\rule[#1pt]{#2pt}{#2pt}\hskip-#2pt
\rule{#2pt}{#1pt}}

\newcommand{\Yfund}{\raisebox{-.5pt}{\drawsquare{6.5}{0.4}}}
\newcommand{\Yasymm}{\raisebox{-3.5pt}{\drawsquare{6.5}{0.4}}\hskip-6.9pt%
        \raisebox{3pt}{\drawsquare{6.5}{0.4}}}

\newcommand{\jref}[4]{{\it #1} {\bf #2}, #3 (#4)}

\newcommand{\NPB}[3]{\jref{Nucl.\ Phys.}{B#1}{#2}{#3}}

\newcommand{\PLB}[3]{\jref{Phys.\ Lett.}{#1B}{#2}{#3}}

\newcommand{\PRD}[3]{\jref{Phys.\ Rev.}{D#1}{#2}{#3}}

\newcommand{\PRL}[3]{\jref{Phys.\ Rev.\ Lett.}{#1}{#2}{#3}}

\newcommand{\PTP}[3]{\jref{Prog.\ Theor.\ Phys.}{#1}{#2}{#3}}

\begin{document}

\begin{titlepage}
\begin{center}
{\hbox to\hsize{hep-th/9703159 \hfill  MIT-CTP-2621}}

\bigskip
\bigskip
\bigskip
             
{\Large \bf  Dynamical Supersymmetry Breaking} \\

\bigskip
\bigskip
\bigskip

{\bf Witold Skiba}\\

\bigskip

{ \small \it Center for Theoretical Physics

Laboratory for Nuclear Science and Department of Physics

Massachusetts Institute of Technology

Cambridge, MA 02139, USA }

{\tt skiba@mit.edu}

\vspace{2cm}
{\bf Abstract}\\
\end{center}

\bigskip
\bigskip

We review mechanisms of dynamical supersymmetry breaking.
Several observations that narrow the search for possible
models of dynamical supersymmetry breaking are summarized.
These observations include the necessary and sufficient
conditions for supersymmetry breaking. The two conditions
are based on non-rigorous arguments, and we show examples
where they are too restrictive. Dynamical effects present
in models with product gauge groups are given special
attention. 

\bigskip

\end{titlepage}

\section{Introduction}
Supersymmetry breaking is an inherent part of any realistic supersymmetric
theory. One of the motivations for supersymmetry is that it stabilizes the
ratio of the electroweak scale to the Planck scale against large radiative
corrections. If supersymmetry is broken dynamically, logarithmic running of
a gauge coupling would also provide an explanation for the smallness of
the ratio of the electroweak and Planck scales.

Almost all phenomenologically viable models consist of two sectors: the
Minimal Supersymmetric Standard Model (MSSM) and the supersymmetry
breaking sector. The information about supersymmetry breaking is transmitted
to the MSSM either by gravitational or by gauge interactions, or a combination
thereof. In the case of gravity mediated supersymmetry breaking, the
electroweak scale, $M_{weak}$, is proportional to
$ \frac{\Lambda_{SB}^2}{M_{Planck}}$, where $\Lambda_{SB}$ is the scale of
supersymmetry breaking and $M_{Planck}$ is the Planck scale.
When supersymmetry breaking is mediated by gauge interactions, $M_{weak}
\propto \alpha^n \Lambda_{SB}$, where $\alpha$ is the structure constant of
the relevant gauge group. In any scenario, the electroweak scale is tied
to the supersymmetry breaking scale. If the scale of supersymmetry breaking
is naturally small, so is the electroweak scale.

The scale of supersymmetry breaking can be small naturally because of the
supersymmetric non-renormalization theorem~\cite{Witten1}. Since the
superpotential receives no radiative corrections, if supersymmetry
is unbroken at the tree-level, it remains unbroken at any order of
perturbation theory. Therefore, only non-perturbative effects
can be responsible for dynamical supersymmetry breaking. 

Consequently, analyzing models of dynamical supersymmetry breaking (DSB)
requires knowledge of the non-perturbative behavior of supersymmetric gauge
theories.  Models of DSB have been around for over ten
years~\cite{ADSprl,ADSplb,ADSnpb}. These models relied on knowledge of
only a small number of dynamical phenomena present in supersymmetric theories,
or sometimes on plausibility arguments when one did not know the dynamics.
After the giant leap in understanding of the low-energy behavior of
supersymmetric theories~\cite{Seiberg}, several new mechanisms and models
of dynamical supersymmetry breaking have been found. The older models
of DSB were analyzed again and there is now a lot more evidence for
supersymmetry breaking in these theories.

In the next section, we discuss general conditions that help to identify
potential candidate theories for DSB\@. Then, in Section~\ref{sec:QCD} we
give a summary of the results on the low-energy dynamics
of supersymmetric QCD\@. We will also illustrate mechanisms of DSB in several
simple examples in that section. In Section \ref{sec:product}, we will
explain how these mechanisms give rise to supersymmetry breaking in more
complicated theories, like the ones with two non-abelian gauge groups.
Afterwards in Section~\ref{sec:exceptions}, we show examples when our general
conditions for supersymmetry breaking do not need to be satisfied.

\section{Some general arguments}
\label{sec:general}

In this section, we summarize both the necessary and sufficient conditions for dynamical supersymmetry breaking. It is necessary for the theories to be
chiral to break supersymmetry dynamically~\cite{Witten2}. It is sufficient
for supersymmetry breaking that a theory without flat directions
has a spontaneously broken global symmetry~\cite{ADSplb,ADSnpb}.
We will also review an observation by Dine, Nelson, Nir and
Shirman~\cite{DNNS}.

The Witten index ${\rm Tr} (-1)^F$ measures the number of bosonic states
of zero energy minus the number of fermionic ones. Since unbroken
supersymmetry implies that the vacuum energy is zero, the Witten index counts
the difference between the number of supersymmetric bosonic and fermionic
vacua. If the index is nonzero, then there are certainly supersymmetric
vacua, so supersymmetry is preserved in the ground state. It turns out
that pure Yang-Mills theories have a non-zero index~\cite{Witten2}.

The index does not change when the parameters of a theory vary continuously.
If it is possible to write the mass terms for all matter fields
in the theory, then all mass parameters can be adjusted to take large
values. Consequently, all matter fields can be decoupled from the theory.
The low-energy theory is pure Yang-Mills, which cannot break supersymmetry.
We will therefore consider chiral theories as candidates for DSB\@. However,
the index can change discontinuously when a change of parameters
alters the asymptotic behavior of the theory. We will return to this
possibility in Section~\ref{sec:exceptions}.

We now turn to the second criterion for DSB\@. A generic supersymmetric gauge
theory without tree-level superpotential has a large set of possible vacuum
states. These are the points where the D-terms vanish, the so-called
flat directions. The D-terms are
\begin{equation}
{\rm D}^{\alpha}=\sum_{i} \Phi^\dagger_i T^{\alpha} \Phi_i,
\end{equation}
where $T^\alpha$ denotes the gauge generators in the representations
under which the chiral superfields $\Phi_i$ transform. Knowledge
of the D-flat directions (or the classical moduli space) is the first step
in analyzing any theory. Finding all values of $\Phi_i$ that satisfy the
equation ${\rm D}^{\alpha} \equiv 0$ for all $\alpha$'s is generally quite
complicated. Some useful techniques were presented in
Refs.~\cite{ADSprl,ADSnpb}. This difficult algebraic exercise can be
circumvented by using a theorem on a one-to-one correspondence between
flat directions and vacuum expectation values of gauge-invariant
holomorphic polynomials~\cite{Markus}. Instead of solving  ${\rm D}^{\alpha}
\equiv 0$, one can find all independent gauge-invariant polynomials 
constructed from chiral superfields. Vacuum expectation values of the
gauge-invariant polynomials parameterize the classical moduli space.
The only difficulty in this approach is ensuring that one has a complete
set of gauge invariants.

When the superpotential is added, some flat directions are lifted, meaning
that the F-terms are usually non-zero along D-flat directions.
It is crucial to know if all flat directions are lifted. The following
procedure is useful. First, compute all F-terms by taking the
equations of motion. Next, make the F-terms gauge invariant by all possible
contractions with chiral superfields. If it is possible to determine
all gauge invariants by setting the F-terms to zero, then the F-terms
vanish only at one point of the moduli space (usually the origin) and
flat directions are lifted~\cite{Markus}.

Let us consider an example: a theory with an $SU(3) \times SU(2)$ gauge
group~\cite{ADSnpb}. The field content of the theory is 
\begin{equation}
  \begin{array}{c|cc}
    & SU(3) & SU(2) \\ \hline 
    Q & \Yfund & \Yfund \\
    \bar{U} & \overline{\Yfund} & 1 \\
    \bar{D} & \overline{\Yfund} & 1 \\
    L & 1 & \Yfund 
  \end{array} \hspace{2cm}
  \begin{array}{l}
    X=Q \bar{U} L \\ Y=Q \bar{D} L \\ Z=Q \bar{U} \, Q \bar{D}
  \end{array}
\end{equation}
where $X$, $Y$ and $Z$ form a complete set of holomorphic gauge invariants.
Take as superpotential
\begin{equation}
\label{eq:W32}
  W=Q \bar{U} L.
\end{equation}
By setting the $\bar{U}$ equation of motion to zero and multiplying
by $\bar{U}$ and $\bar{D}$ one obtains $X=0$ and $Y=0$. Similarly,
the F-term for $L$ sets $Z$ to zero, thus the superpotential of
Eq.~\ref{eq:W32} lifts all flat directions. We will return to this theory
in the next section and describe its quantum mechanical behavior.

We are now ready to present the sufficient condition for supersymmetry
breaking. Suppose a theory does not have flat directions either because
it does not have any gauge invariants constructed from chiral superfields
or because flat directions are lifted by appropriate choice of tree-level
superpotential. If such a theory has a continuous global symmetry
which is spontaneously broken, then supersymmetry
is also spontaneously broken~\cite{ADSplb,ADSnpb}. A spontaneously broken
global symmetry implies the presence of a Goldstone boson. In a
supersymmetric theory, this Goldstone boson has to combine with another
massless scalar particle to form a supersymmetric multiplet. Since
we assume that there are no non-compact flat directions, then there cannot
be another massless scalar in the theory. Hence, supersymmetry must be broken.
We want to stress that this argument applies to any global symmetry, not
necessarily an R-symmetry. R-symmetry is especially useful when the
superpotential is a generic function of chiral superfields consistent
with symmetries~\cite{NS}. 

The program looks quite simple: take a chiral theory, then lift all flat
directions without explicitly breaking all global symmetries. This,
however, is more complicated than it seems. An interesting observation
by Dine, Nelson, Nir and Shirman helps to find such theories.
Suppose one knows a theory that breaks supersymmetry dynamically.
Instead of the original theory consider a theory with the gauge
group reduced to a subgroup. The matter fields are the same, except
that the representations they transform in are obtained by decomposing
the original representations into those of the subgroup. Such a theory
is guaranteed to be anomaly-free just as the the original one was.
Moreover, it is frequently possible to lift all flat directions
while preserving  a global symmetry. We should stress here that we do not
imply a physical procedure of breaking the gauge group by the Higgs
mechanism. Also, the superpotential in the new theory is not derived
from the original one. The theory with reduced gauge group has fewer
D-terms, while the same number of chiral superfields. Usually, lifting flat
directions in the new theory requires additional terms in the tree-level
superpotential.

It is easy to understand why this procedure is likely to yield
new theories of DSB\@. Suppose one adds an adjoint chiral superfield to the
theory. This field in the vector-like representation does not affect the
Witten index. The gauge symmetry can be broken by arranging the
superpotential for the adjoint, where gauge bosons corresponding to broken
symmetries and uneaten fields from the adjoint become massive. This idea
is quite difficult to carry out explicitly in general~\cite{LRR}.  
Depending on the chosen pattern of gauge symmetry breaking, the adjoint
may not be sufficient, and one needs other fields in vector-like
representations to achieve the breaking.

\section{Basic mechanisms for DSB}
\label{sec:QCD}
The obvious question to be addressed next is what kind of non-perturbative
phenomena can be responsible for spontaneous global symmetry breaking and
as a result supersymmetry breaking. We first review the low-energy effects
in supersymmetric QCD, and then illustrate how the non-perturbative dynamics
can lead to DSB\@. Refs.~\cite{Seiberg,lectures} contain a detailed review
about supersymmetric QCD and analysis of other theories, as well as
extensive collection of references.

Supersymmetric QCD is an $SU(N_c)$ theory with $N_f$ fields $Q^i$ in the
fundamental representation and $N_f$ fields $\bar{Q}_j$ in the antifundamental.
The indices $i,j=1,\ldots,N_f$ denote flavor degrees of freedom, while the
color indices are suppressed. Classically, the flat directions are
parameterized by ``meson'' fields $M_i^j=\bar{Q}_i Q^j$, and ``baryon'' fields
$B_{i_{N_c+1} \ldots i_{N_f}}=\epsilon_{i_1 \ldots i_{N_f}} Q^{i_1} \ldots
Q^{i_{N_c}}$, $\bar{B}^{i_{N_c+1} \ldots i_{N_f}}=
\epsilon^{i_1 \ldots i_{N_f}} \bar{Q}_{i_1} \ldots \bar{Q}_{i_{N_c}}$.
Baryon fields exist only when $N_f \geq N_c$. The mesons and baryons are
not independent. They obey constraints
\begin{equation}
 \begin{array}{c}
   B_{i_{1} \ldots i_{N_f-N_c}} M^{i_1}_j=0 ,\; \;
  \bar{B}^{i_{1} \ldots i_{N_f-N_c}} M_{i_1}^j=0 ,\; \\
  M^{i_1}_{j_1} \ldots M^{i_{N_c}}_{j_{N_c}} \epsilon_{i_1 \ldots i_{N_f}}
  \epsilon^{j_1 \ldots j_{N_f}} = B_{i_{N_c+1} \ldots i_{N_f}}
  \bar{B}^{i_{N_c+1} \ldots i_{N_f}}.
 \end{array}
\end{equation}
These constraints are easy to verify if we express them in terms of the
underlying fields $Q^i$ and $\bar{Q}_i$. Vacuum expectation values (VEVs)
of these gauge invariant polynomials, subject to constraints, describe
the classical moduli space. The classical theory has a large set of
degenerate vacuum states. For many of these states, the degeneracy is
accidental---not protected by symmetries---and can be removed by quantum
effects. The quantum-mechanical picture depends on the number of flavors
$N_f$~\cite{Seiberg}. We explain briefly each interesting case.

When $N_f<N_c$, non-perturbative effects generate a superpotential 
of the form
\begin{equation}
  W_{dyn}=\left( \frac{\Lambda^{3 N_c-N_f}}{\det M} \right)^{1/(N_c-N_f)}.
\end{equation}
The scalar potential corresponding to this superpotential is inversely
proportional to the VEVs of chiral superfields. Therefore, supersymmetric QCD
has no stable vacuum state for $0<N_f<N_c$. It is possible to lift flat
directions by adding small mass terms and have
a stable vacuum state. In the ground state all fields have VEVs, and the
gauge group is broken to $SU(N_c-N_f)$. Even though flat directions are lifted
by adding mass terms, supersymmetric QCD does not break supersymmetry since
it is non-chiral\footnote{The $SU(N_f)\times U(1)_B$ global symmetry
preserved by the mass terms is not broken in the ground state.}.

The low-energy theory with $N_f=N_c$ confines. The physical degrees of freedom
are $N_f^2$ mesons $M^i_j$ and baryons $B$, $\bar{B}$. Classically, these
fields obey the constraint $\det M-B \bar{B}=0$. In the quantum regime,
the constraint is modified: $\det M-B \bar{B}=\Lambda^{2 N_f}$, so the
classical and quantum moduli spaces are different. It is important that the
origin of the moduli space---where all fields have zero expectation
values---does not belong to the quantum moduli space. Quantum mechanically,
some of the fields necessarily have VEVs; consequently some global symmetries
are always broken. The constraint can be implemented by including it with
a Lagrange multiplier $\mu$ in the superpotential:
\begin{equation}
  W=\mu (\det M-B \bar{B} - \Lambda^{2 N_f}).
\end{equation}

A theory with one more flavor $N_f=N_c+1$ is also confining. In this case,
the classical and quantum moduli spaces are identical. The gauge invariant
operators $M$'s, $B$'s and $\bar{B}$'s correspond to physical degrees of
freedom describing the theory at the origin. The mesons and baryons interact
via a ``confining superpotential''
\begin{equation}
  W=\frac{1}{\Lambda^{2 N_c-1}} \left(\bar{B}^i M_i^j B_j -\det M \right).
\end{equation}
One of the consistency checks on the confining picture is that 't~Hooft anomaly
matching conditions between the high and low-energy degrees of freedom are
satisfied. In the previous confining case, $N_f=N_c$, anomalies do not match
at the origin, which indicates that the origin is not part of the moduli
space. However, anomalies match on the points that belong to the quantum
moduli space.

For a larger number of flavors $N_c < N_f < 3 N_c$, supersymmetric QCD
is either in the free-magnetic or conformal phase. Its infrared fixed  point
can be described equivalently in terms of another theory---supersymmetric
QCD with $N_f-N_c$ colors. The ``dual'' theory has also $N_f$ flavors of
``magnetic quarks'' $q_i$, $\bar{q}^i$; $N_f^2$ elementary gauge singlet
``mesons'' $\tilde{M}^i_j$ and the following
superpotential
\begin{equation}
  W= \tilde{M}^i_j \bar{q}^j q_i.
\end{equation}
Gauge invariant operators of the original $SU(N_c)$ correspond to the
gauge invariants of the dual $SU(N_f-N_c)$. The mesons $\bar{Q} Q$ are
mapped into gauge singlet fields $\tilde{M}$, while baryon operators
$Q^{N_c}$ are mapped into baryons $q^{N_f-N_c}$.

Supersymmetric QCD is infrared free for $N_f\geq 3 N_c$ and has no interesting
non-perturbative dynamics. We again refer the reader to the original
papers~\cite{Seiberg} and the lecture notes~\cite{lectures} for more details.

Let us now make use of these results and observations from
Section~\ref{sec:general}. The first mechanism which can break supersymmetry
is the dynamically generated superpotential. We will analyze in detail
the well-known 3-2 model~\cite{ADSnpb}, whose field content we already
presented in the previous section. The model has two gauge groups: $SU(3)$
and $SU(2)$. Let us first assume that the dynamical scale of the $SU(3)$
interactions, $\Lambda_3$, is much larger than that of $SU(2)$. In this limit,
non-perturbative effects of the $SU(2)$ dynamics can be neglected.
$SU(2)$ gauge group makes the theory chiral and also lifts some
flat directions. The $SU(3)$ theory is an example of supersymmetric
QCD with $N_f < N_c$. It generates a dynamical superpotential
$W_{dyn}=\frac{\Lambda^7_3}{Q\bar{U}\, Q\bar{D}}$.

We have seen that the tree-level superpotential $W=Q \bar{U} L$ lifts all flat
directions, it also preserves a $U(1)_R$ symmetry. Because of the dynamically
generated superpotential, some fields get VEVs which break the R-symmetry.
Thus, supersymmetry must be broken as well. One can check that the
vacuum energy is non-zero when the F-terms are computed from the full
superpotential
\begin{equation}
  W=\frac{\Lambda^7_3}{Q\bar{U}\, Q\bar{D}} + Q \bar{U} L.
\end{equation}

What happens in the 3-2 model when the $SU(2)$ dynamics is more important,
that is when $\Lambda_2 \gg \Lambda_3$? The $SU(2)$ has four doublets, so it
has a quantum modified constraint. We also expect supersymmetry to be broken.
All global symmetries are preserved only at the origin of the moduli space.
Since in the quantum theory the origin does not belong to the moduli,
some global symmetries and also supersymmetry can be broken~\cite{IT1}.
Again, one can explicitly check that the full superpotential
\begin{equation}
  W=\mu \left[{\rm Pf} (Q L) (Q Q)  -\Lambda_2^4 \right] + (Q \bar{U} L)
\end{equation}
breaks supersymmetry. We indicated degrees of freedom confined by the
$SU(2)$ dynamics in parenthesis, they are the physical fields.
The description of the 3-2 model can be found for arbitrary ratio
$\Lambda_3 /\Lambda_2$~\cite{IT1}. The model breaks supersymmetry
in the whole region of parameter space, as expected from arguments about
spontaneously broken global symmetries in the absence of flat directions.

Not only the dynamically generated superpotential or the quantum modified
constraint that can lead to DSB\@. Confinement can also cause supersymmetry
breaking. When a theory confines the physical degrees of freedom
are gauge-invariant fields. The low-energy description of the theory has to
be written in terms of the confined fields. If the theory has a tree-level
superpotential added to lift flat directions, after confinement the
superpotential has to be re-expressed in terms of the new physical fields.
Because of that, the low-energy theory can have the form of an
O'Raifeartaigh model.

Let us consider an $SU(2)$ theory with one field $Q$ in the three-index
symmetric representation of $SU(2)$~\cite{ISS}. The authors of Ref.~\cite{ISS}
argued that this theory confines at low energies and there is one light
confined field $T=Q^4$. Global symmetries of the theory do not allow a
dynamically generated superpotential. The tree-level superpotential
$W=Q^4$ lifts the flat directions. In the infrared, this superpotential
should be written as $W=T$, which breaks supersymmetry\footnote{In fact,
since $T$ is a free field this model has an accidental $U(1)$ symmetry. A
linear combination of this accidental symmetry and the R-symmetry is preserved
by superpotential. If $T$ has non-zero VEVs at the ground state, the
$U(1)$ symmetry is spontaneously broken.}. It is important that the K\"ahler
potential for the $T$ field is not singular. Classically, the K\"ahler
potential expressed as a function of the invariant $Q^4$ has a singularity
at the origin.

Another example is perhaps more illustrative. We consider an $SU(7)$ theory
with two antisymmetric tensors $A^i$ and six antifundamentals
$\bar{Q}_a$~\cite{us1}. Both $i=1,2$ and $a=1,\ldots,6$ are flavor indices.
This theory behaves like supersymmetric QCD with $N_f=N_c+1$; it confines
and has a confining superpotential. There are two kinds of confined degrees
of freedom: $H =A \bar{Q}^2$ and $N=A^4 \bar{Q}$,
in terms of which the confining superpotential is $W=\frac{1}{\Lambda^{13}}
(N)^2 (H)^2 $. This theory is chiral and it is possible to lift all the
flat directions with the following superpotential
\begin{equation}
  W=
A^1 \bar{Q}_1 \bar{Q}_2 + A^1 \bar{Q}_3 \bar{Q}_4 + A^1 \bar{Q}_5 \bar{Q}_6 +
A^2 \bar{Q}_2 \bar{Q}_3 + A^2 \bar{Q}_4 \bar{Q}_5 + A^2 \bar{Q}_6 \bar{Q}_1.
\end{equation}
After confinement, the tree-level terms turn into linear terms, and
the full superpotential is
\begin{equation}
W=H^1_{12}+H^1_{34}+H^1_{56}+H^2_{23}+H^2_{45}+H^2_{61}+
\frac{1}{\Lambda^{13}} (N)^2 (H)^2.
\end{equation}
Since at low energies fields $H$ and $N$ are to be interpreted as elementary
degrees of freedom, this theory takes the form of an O'Raifeartaigh model.
The tree-level linear terms force VEVs for some fields. These VEVs break global
symmetries and supersymmetry turns out to be broken as well~\cite{us1}.

Another interesting example of theory that breaks supersymmetry because of
strong dynamics is an $SU(5)$ theory with one antisymmetric tensor and
one antifundamental~\cite{ADSplb}. This theory has no flat directions since
there are no invariants that can be constructed out of single ${\bf 10}$ and
single ${\bf \bar{5}}$ of $SU(5)$. This theory has two $U(1)$ symmetries.
The authors of Ref.~\cite{ADSplb} argued that one of these symmetries must
be broken in the ground state. Thus, supersymmetry is broken as well since the
theory has no flat directions. The theory has been analyzed recently
by adding fields in vector-like representations. It is then possible to find
low-energy description of these $SU(5)$ theories with extra fields. Theories
with mass terms for the additional fields break supersymmetry~\cite{PTP}.
A similar theory is $SO(10)$ with one spinor field. It also does not have
flat directions and breaks supersymmetry~\cite{ADSplb}. This model has also
been studied with larger matter content and appropriate mass terms~\cite{MPS}.
 
Note that the 3-2 model can be obtained from the $SU(5)$ theory with ${\bf 10}$
and ${\bf \bar{5}}$ by decomposing $SU(5)$ into its $SU(3)\times SU(2)$
subgroup. This is an illustration of the observation by Dine, Nelson, Nir
and Shirman explained in the previous section. Other possible decompositions
of the $SU(5)$ theory like $SU(4)\times U(1)$~\cite{DNNS} or
$Sp(4)\times U(1)$~\cite{us1} also break supersymmetry. It is interesting
that these theories are chiral only because of the $U(1)$ charge assignment.

In the next section we will analyze theories whose field content is obtained
by decomposing an $SU(N)$ theory with an antisymmetric tensor $A$ and $N-4$
antifundamentals $\bar{F}_i$, $N$ is odd and larger than 5. Let us outline
the mechanism of supersymmetry breaking in the $SU(N)$ theory with an
antisymmetric tensor and $N-4$ antifundamentals~\cite{ADSnpb}.
Without tree-level superpotential this theory has flat
directions described by the gauge invariants $A\bar{F}_i\bar{F}_j $. Along
a generic flat direction, the $SU(N)$ gauge group is broken to $SU(5)$.
The uneaten fields are ${\bf 10}$ and ${\bf \bar{5}}$ of $SU(5)$. The vacuum
energy in the $SU(5)$ theory is proportional to the dynamical scale of
$SU(5)$: $E_{vac} \propto \Lambda_5$.

When $SU(N)$ is  broken to $SU(5)$ by VEVs of order $\langle \phi \rangle$
the scales of $SU(N)$ and $SU(5)$ are related by matching:
\begin{equation}
  \Lambda_5=\Lambda_N^{(2N+3)/13} \, \langle \phi \rangle^{- (2 N-10)/13}.
\end{equation}
Here, $\langle \phi \rangle$ indicates a generic value of a VEV for either
$A$ or $\bar{F}$~\footnote{These VEVs are related because of the D-flatness
conditions.}. Therefore, the vacuum energy as a function of the VEVs is
$E_{vac} \propto \langle \phi \rangle^{- (2 N-10)/13}$. This resembles
the situation in models with a dynamically generated superpotentials.
The low-energy $SU(5)$ generates a potential which decreases to zero at large
VEVs. When flat directions are lifted by the tree-level superpotential
$W=\lambda^{ij} A \bar{F}_i \bar{F}_j$, the theory breaks
supersymmetry~\cite{ADSnpb}. Here, $\lambda^{ij}$ is a matrix of rank $N-5$.

\section{Product groups}
\label{sec:product}
In this section we describe models obtained by decomposing the field content
of the $SU(N)$ theory with an antisymmetric tensor and $N-4$ antifundamentals
into $SU(N-M)\times SU(M)\times U(1)$ subgroup. Depending on $N$ and $M$,
the two gauge groups have different number of flavors. By adjusting $N$ and
$M$ we can analyze theories in different phases.

$SU(N)$ theories with an antisymmetric tensor, $N-4$ antifundamentals and
some number of fundamental-antifundamental pairs behave very much like
supersymmetric QCD~\cite{PTP}. When $N_f < 3$, the theory generates a dynamical
superpotential. (A superpotential is not generated when $N_f=0$ and N is odd,
since all holomorphic gauge invariants vanish classically.) For $N_f=3,4$
the theory confines respectively with a quantum deformed moduli space or
with a confining superpotential. When the number
of flavors is larger than four the theory admits a dual description.

The fields of $SU(N)$ decompose under $SU(N-M)\times SU(M) \times U(1)$
as follows
\begin{eqnarray}
 A & \rightarrow & A(\Yasymm,1)_{2 M} + a(1,\Yasymm)_{2 M-2 N} +
 T(\Yfund,\Yfund)_{2 M-N}, \nonumber \\ 
 \bar{F}_i & \rightarrow & \bar{F}_i(\overline{\Yfund},1)_{-M} + 
                         \bar{Q}_i(1,\overline{\Yfund})_{N-M},
\end{eqnarray}
where the subscripts indicate the $U(1)$ charge. It is quite tedious to
show that the following superpotential lifts all flat directions and
preserves an R-symmetry~\footnote{This R-symmetry is anomalous with
respect to the $U(1)$ gauge group. However, the Goldstone boson resulting
from the spontaneous breaking of this symmetry is massless.
Therefore, the argument for supersymmetry breaking when a global
symmetry is broken still holds.}
\begin{eqnarray}
\label{eq:Wtree}
W_{tree}& = & A \bar{F}_1 \bar{F}_2 + \ldots + A \bar{F}_{N-6} \bar{F}_{N-5} +
            a \bar{Q}_2 \bar{Q}_3 + \ldots + a \bar{Q}_{N-5} \bar{F}_{1} + 
         \nonumber \\ 
        & & T \bar{F}_1 \bar{Q}_1 + \ldots + T \bar{F}_{N-4} \bar{Q}_{N-4}. 
\end{eqnarray}
The field $T$ is the only field, which transforms under both gauge groups.
Through that field the two groups can affect each other's dynamics.

Models obtained by decomposing $SU(N)$ to $SU(N-1)\times U(1)$ and
$SU(N-2)\times SU(2) \times U(1)$ were analyzed in Ref.~\cite{DNNS}.
In both cases the $SU(N-1)$ or $SU(N-2)$ groups generate a dynamical
superpotential. Not surprisingly, these theories break supersymmetry when flat
directions are lifted~\cite{DNNS}. Let us examine in detail the decomposition
of $SU(7)$ to $SU(4)\times SU(3) \times U(1)$~\cite{us2}. When the gauge groups
are analyzed independently, $SU(4)$ has an antisymmetric tensor and three
flavors. It confines with a quantum modified constraint. Because the
modified constraint breaks the R-symmetry, one expects supersymmetry to
be broken in the limit $\Lambda_4 \gg \Lambda_3$.

We will analyze the theory in the opposite limit $\Lambda_4 \ll \Lambda_3$
and show that supersymmetry is broken as well. Above the scale $\Lambda_4$,
$SU(4)$ is still weakly gauged and we can neglect its non-perturbative
dynamics. Below $\Lambda_3$, the $SU(3)$ confines since it is a supersymmetric
QCD with $N_f=N_c+1$. We have to describe the physics in terms of confined
mesons and baryons. These are
\begin{equation}
    B=(T^3),\; \; 2\cdot \bar{B}=(a\bar{Q}^2),\; \;
      \bar{B}=(\bar{Q}^3), \; \;
    3\cdot M=(T\bar{Q}),\; \; M=(T a)
\end{equation}
Note that in the case of $SU(3)$ the field $a$ transforms as an
antifundamental, just like the $\bar{Q}$'s. Since the underlying fields
transform under $SU(4)$, some of the composites objects carry $SU(4)$
quantum numbers. In particular, $B$ transforms as an antifundamental of
$SU(4)$, $M$'s as fundamentals while $\bar{B}$'s are singlets.

After confinement of $SU(3)$, the field content of the $SU(4)$ group has
changed. The $SU(4)$ has now one more flavor. Above the $SU(3)$ confining
scale, the $SU(4)$ gauge group has three flavors in addition to an
antisymmetric tensor. Below $\Lambda_3$, there are four flavors:
four fundamentals $M$ and four antifundamentals $\bar{F}_i$ and $B$.
The effective $SU(4)$ theory below $\Lambda_3$ is an analog of supersymmetric
QCD with $N_f=N_c+1$ and not $N_f=N_c$. It confines and has a confining
superpotential. When we go below the new effective scale $\Lambda'_4$,
we have to again change the description into gauge invariant fields.
It turns out that by simply adding the confining superpotentials obtained
from $SU(3)$ and $SU(4)$ gauge dynamics one obtains a correct description
of the theory for any ratio $\Lambda_3 /\Lambda_4$~\cite{us2}. For instance,
one of the equations of motion properly reproduces the quantum modified
constraint expected in the $\Lambda_4 \gg \Lambda_3$ limit.
When the tree-level superpotential of Eq.~\ref{eq:Wtree} is added
\begin{equation}
  W=W_{conf}^{SU(3)} + W_{conf}^{SU(4)} + W_{tree},
\end{equation}
the low-energy theory is an O'Raifeartaigh model and breaks supersymmetry.
Again, the tree-level terms become linear and force VEVs for some fields.
The $U(1)$ gauge group does not play any dynamical role, its sole
purpose is to lift some flat directions. A model with similar dynamics based
on the $SU(3) \times SU(2)$ gauge group was analyzed in Ref.~\cite{IT1}.

In the above example we saw that the dynamics of one gauge group affected
dynamics of the other---it changed  the effective number of flavors. There is,
however, another way to see that supersymmetry is broken in the $\Lambda_4 \ll
\Lambda_3$ limit, which will be essential for other examples. So far, we have
first analyzed the theory without including the tree-level superpotential.
We found the low-energy description without superpotential, then added
the superpotential and checked if supersymmetry is broken. The tree-level
superpotential may play an important dynamical role, it can alter the
infrared behavior of the theory.

Suppose we included the superpotential of Eq.~\ref{eq:Wtree} from the
beginning. We again study the theory in the limit $\Lambda_4 \ll \Lambda_3$.
After confinement in $SU(3)$, the superpotential has to be expressed in terms
of $SU(3)$ invariants. In particular, all terms of the form
$T\bar{F}_i \bar{Q}_i$ will become mass terms
$M \bar{F}$, where $M=( T\bar{Q})$ transforms as a fundamental of $SU(4)$.
The mass parameters associated with these terms are $\lambda \Lambda_3$,
where $\lambda$ is a Yukawa coupling. For brevity, Yukawa couplings were
not explicitly specified in Eq.~\ref{eq:Wtree}. If $\lambda \Lambda_3 >
\Lambda_4$, we need to integrate out the massive fields before we take
into account the $SU(4)$ dynamics. In the tree-level superpotential
there are mass terms for three flavors
of $SU(4)$. After integrating them out we obtain an $SU(4)$ theory with
an antisymmetric tensor and one flavor, which generates a dynamical
superpotential. The dynamically generated superpotential is the reason
for supersymmetry breaking in this regime. Of course, this limit could
have been recovered from the exact description of the theory
without superpotential. It is sometimes easier to analyze the theory
taking the tree-level superpotential into account.

We now consider another example obtained by decomposing $SU(N+4)$ theory
with an antisymmetric tensor and $N$ fundamentals into $SU(N)\times SU(4)
\times U(1)$ subgroup, $N \geq 5$~\cite{us2}. When the two non-abelian
gauge groups are examined independently, the $SU(N)$ is an analog of
supersymmetric QCD with $N_f=N_c+1$, it confines without modifying the moduli
space. $SU(4)$ has an antisymmetric tensor and $N$ flavors. Depending on $N$,
$SU(4)$ can be in the free-magnetic, conformal or even infrared-free
phase. Naively, there is no mechanism that would cause supersymmetry
breaking. It seems that the theory could have a vacuum state at the origin,
where no symmetries are broken and the tree-level superpotential vanishes.

We will investigate the theory in the limit $\Lambda_N \gg \Lambda_4$, and
take the superpotential of Eq.~\ref{eq:Wtree} into account from the start.
The analysis is in fact quite similar to the one of the $SU(4)\times SU(3)
\times U(1)$ example. When $SU(N)$ confines, the $T\bar{F}_i \bar{Q}_i$ terms
become mass terms $M_i \bar{Q}_i$. We can integrate out these terms and obtain
an $SU(4)$ theory with an antisymmetric tensor and one flavor. Because
of the dynamically generated superpotential in $SU(4)$, supersymmetry is
broken.

These examples may seem to suggest that it is the confining dynamics that
is crucial for supersymmetry breaking. Let us consider an $SU(N) \times SU(5)
\times U(1)$ theory obtained by decomposing $SU(N+5)$~\cite{us2}. The $SU(N)$
gauge group can be given a dual description. In the dual, meson operators are
mapped into elementary singlet fields. If we include the tree-level
superpotential of Eq.~\ref{eq:Wtree}, all $T\bar{F}\bar{Q}$ terms become mass
terms $\tilde{M} \bar{Q}$. Here, $\tilde{M}$ singlets of the dual gauge group
still carry $SU(5)$ charges, since $SU(5)$ was just a spectator in dualizing
$SU(N)$. In the dual description, the effective number of $SU(5)$ flavors
changes first as a result of duality transformation and second because of
the mass terms from the tree-level superpotential. Effective $SU(5)$ theory
again generates a dynamical superpotential which breaks supersymmetry.

These examples show that supersymmetry can be broken in product group theories
in many cases where one would not expect appropriate dynamical effects to
take place. Dynamics of one gauge group can affect the dynamics of other
group. It is interesting that the tree-level superpotential is more
important than simply lifting flat directions. It can change the phase
of the theory. Many other examples of product group theories and theories
utilizing dual gauge group dynamics were presented in Ref.~\cite{IT2,PTS}.

\section{Other possibilities}
\label{sec:exceptions}

In this section we point out that there can be non-chiral theories which break
supersymmetry, and also that in certain cases it is not necessary to lift all
classical flat directions. We first consider an $SU(2)$ theory with four
doublets $Q_i$ and six singlets $S^{ij}$~\cite{IT1,IzYa}. This theory has a
global $SU(4)$ symmetry under which $Q_i$'s transform as a fundamental while
$S^{ij}$'s as an antisymmetric tensor. The $SU(2)$ has the same number of
flavors as colors, so it has a dynamically modified constraint. The tree-level
superpotential, $W_{tree}=\lambda S^{ij} Q_i Q_j$, preserves the global
$SU(4)$. This superpotential lifts flat directions associated with $Q$'s,
however the singlet fields remain flat. In the quantum theory, the full
superpotential includes the modified constraint in addition to the
tree-level term,
\begin{equation}
  W=\mu ({\rm Pf}\, M_{ij} -\Lambda_2^4) + \lambda S^{ij} M_{ij},
\end{equation}
where $M_{ij}=(Q_i Q_j)$. The equations of motion with respect to
$S^{ij}$ set all $M$'s to zero, which is incompatible with the constraint.
The sufficient condition for supersymmetry breaking is not useful
in this case.  Quantum modified constraint breaks the global $SU(4)$ symmetry,
but the theory has classical flat directions.

This $SU(2)$ theory is not chiral. The Witten index argument fails in this
case because of flat directions associated with the $S$ fields. When a mass
term $m {\rm Pf}\, S$ is added, the asymptotic behavior of potential for the
$S$ fields changes. The form of the potential at infinity changes
discontinuously when $m$ becomes non-zero. In fact, the theory with mass terms
has supersymmetric vacua at expectation values of $S$ proportional to
$\frac{\lambda \Lambda_2^2}{m}$~\cite{IT1}. These supersymmetry-preserving
vacua disappear from the theory when $m$ is zero.

It turns out that classical flat directions associated with $S^{ij}$ are
not flat when quantum corrections are taken into account~\cite{Yuri}.
In the limit where $\lambda \langle S \rangle > \Lambda_2$ one finds that the
vacuum energy is proportional to $\sqrt{\lambda} \Lambda_2$. When the one-loop
running of the Yukawa coupling $\lambda$ is examined, one finds that $\lambda$
as a function of $S$ has a minimum. Therefore, the full scalar potential for
$S$ is not flat when the deviation of the K\"ahler potential from its
canonical form is included.

The $SU(2)$ theory illustrates another important point: supersymmetry can
be broken even when the classical theory has flat directions. This can happen
if along the flat directions the gauge group is not completely broken. In
the $SU(2)$ example, VEVs of the singlet fields obviously do not break the
gauge group. One is usually worried that flat directions may lead to runaway
theories without vacuum state. If the gauge group is completely broken, the
larger the VEVs are, the weaker the dynamical effects become. When there is
an unbroken subgroup, its effective scale can grow with the vacuum expectation
values and lift the flat direction quantum mechanically~\cite{Yuri}.

In the $SU(2)$ theory considered above, when $\lambda \langle S \rangle$ is
large, one can integrate out the doublets and obtain a pure Yang-Mills theory
with a low-energy scale $\Lambda_{2,L}=({\rm Pf} S)^{1/6} \Lambda_2^{2/3}$. The
scale increases with the expectation values of $S^{ij}$'s. The same can happen
in chiral theories.  For example, the superpotential of Eq.~\ref{eq:Wtree}
lifts all  flat directions of $SU(N-M)\times SU(M)\times U(1)$ theories
described in the previous section. The theory with all $a\bar{Q}_i \bar{Q}_j$
terms left out has classical flat directions. Along a generic flat direction
$SU(N-M)$ is unbroken, while $SU(M)$ is completely broken. When fields
$\bar{Q}$ get VEVs, $T\bar{F} \langle \bar{Q} \rangle $ terms are mass terms
for $SU(N-M)$ flavors. There are $M$ such mass terms, since $T$ is at most rank
$M$. After integrating these out, the effective $SU(N-M)$ has an antisymmetric
tensor and $N-M-4$ flavors. This is exactly of the form of the original
theory from which $SU(N-M)\times SU(M)\times U(1)$ models were derived.
Supersymmetry breaking for this theory was described at the end of
Section~\ref{sec:QCD}. The vacuum energy in the effective $SU(N-M)$
is proportional to $\Lambda_{N-M}$, which grows with expectation values
of the $a \bar{Q} \bar{Q}$ flat directions as a result of scale
matching~\cite{Yuri}. Therefore, these flat directions are lifted in the
quantum theory and supersymmetry is broken.
 
\section{Conclusions}
\label{sec:conclusions}

We have described the basic facts about models of dynamical supersymmetry
breaking. Chiral theories without flat directions in which global symmetries
are spontaneously broken also break supersymmetry. The three dynamical
mechanisms that can cause supersymmetry breaking are dynamically generated
superpotential, quantum modified constraint and confinement. The picture
is much more complicated in product group theories. The dynamics of one gauge
group can affect the dynamics of the others, and the interplay among the
dynamics of the product groups can be non-trivial. This interplay, together
with the tree-level superpotentials, can change the effective number of
flavors and yield theories which break supersymmetry, even in cases were
one would not expect supersymmetry breaking. There exist examples of
non-chiral theories or theories with flat directions, which nevertheless
break supersymmetry.

We have focused on the technical aspects of finding theories which break
supersymmetry. However, we have not addressed the most important problem,
that is how to identify which theory will describe nature most accurately.
So far, there is no answer to that question. The solution depends strongly
on the mechanism of communicating supersymmetry breaking to the MSSM\@.
Refs.~\cite{DNNS,PT,communication} contain several recently proposed scenarios
of gauge mediation. One cannot argue for or against a given model on the basis
of simplicity. One may want to ``unify'' the DSB sector with the messenger
sector and MSSM~\cite{PT}. This requires theories with relatively large
gauge groups.

We also have not discussed the properties of the supersymmetry-breaking
ground state in presented models. We were satisfied with showing that the
vacuum energy is not zero in each case. The particle spectrum and the unbroken
symmetries at the ground state are important for communicating supersymmetry
breaking to the visible sector. There are at least two obstacles in identifying
the properties of the ground state. First, in many theories we do not know the
K\"ahler potential. If the theory is weakly coupled, one can calculate the
K\"ahler potential in perturbation theory. In some strongly coupled theories,
one expects the K\"ahler potential to be canonical, up to a constant, near
the origin. However, there is no systematic way of computing the corrections.
The second difficulty is technical. Many of the analyzed theories contain
a large number of fields. Finding the minimum for so many variables, even
numerically, is not an easy task.

Hopefully, with so many recently constructed models of DSB, one of the theories
is the right one. If not, the tools for building models we have already
acquired are perhaps enough to find it.

\section*{Acknowledgments}
It is a pleasure to thank Csaba Cs\'aki, Lisa Randall and Martin Schmaltz
for fruitful collaborations. Many of the results presented here are the
outcome of these collaborations. I am also grateful to Daniel Freedman
and Asad Naqvi for their comments on the manuscript. This work is
supported in part by the U.S. Department of Energy under cooperative
agreement \#DE-FC02-94ER40818. 

\end{document}